\documentclass[12pt]{article}
\usepackage{graphicx}
\usepackage[bookmarksnumbered,colorlinks,plainpages]{hyperref}
\usepackage{amssymb,amsfonts,amsmath}
\usepackage[T2A]{fontenc}

\begin{document}

\title{Least action principle and stochastic motion : a generic derivation of path probability}
\author{Aziz El Kaabouchi$^1$ and Qiuping A. Wang$^{1,2}$
\thanks{Email: awang@ismans.fr} \\
{\small $^1$LP2SC, ISMANS, LUNAM Universit\'e}, \\ {\small 44, Avenue, F.A. Bartholdi, 72000, Le Mans, France.} \\
{\small $^2$IMMM, UMR CNRS 6283, Universit\'e du Maine, 72085 Le Mans, France}}

\date{}

\maketitle

\begin{abstract}
This work is an analytical calculation of the path probability for random dynamics of mechanical system described by Langevin equation with Gaussian noise. The result shows an exponential dependence of the probability on the action. In the case of non dissipative limit, the action is the usual one in mechanics in accordance with the previous result of numerical simulation of random motion. In the case of dissipative motion, the action in the exponent of the exponential probability is just the one proposed in a previous work (Q.A. Wang, R. Wang, arXiv:1201.6309), an action defined for the total system including the moving system and its environment receiving the dissipated energy. In both cases, the result implies that the most probable paths are the paths of least action which, in the limit of vanishing randomness, become the regular paths minimizing the action.
\end{abstract}

PACS numbers: 

\newpage

\section{Introduction}

A characteristic of random motion is the existence of many possible paths between two given positions. A trivial experiment to show this is to drop a small feather in the air from a top point to a fixed point below. The first time it goes down along a zigzag path, the second time another path, and so on. If each path is defined as a band with a reasonable thickness instead of a single geometrical line, it is possible to assign a probability of occurrence to each of these paths, or the probability for a particle to take. In principle, this probability can be experimentally measured by observing a large number of identical feathers moving down between the two points, and estimated by the number of feathers observed along this path, divided by the total number of feathers arriving at the end point below. In practice, this kind of experiments is not that easy because, in order to get reliable calculation of probability, one has to count a very large number of identical bodies moving between two points under identical conditions.

In a recent work\cite{Wang1}, we have proposed numerical measure of this probability by simulating the random motions subject to conservative forces. The aim of this probability measure is to determine its random variables. The random motion was generated by a Langevin equation with Gaussian noise. This numerical model is different from the above example of feathers in that there is no friction or energy dissipation on the falling bodies. This is an ideal motion or an asymptotic model for weakly damped random motion. The choice of this model, fully discussed in \cite{Wang1}, is motivated by a theoretical study of path probability in the framework of a probabilistic Hamiltonian mechanics based on a generalization of least action principle of the regular Hamiltonian mechanics\cite{Wang2,Wang3,Wang4}. According to this theory, if the system is perturbed by a Gaussian white noise and statistically remains Hamiltonian one without or with very weak dissipative force, the path probability should be exponentially decreasing function $e^{-\gamma A}$ of the action $A=\int (K-V)dt$ where $K$ is the kinetic energy, $V$ the potential energy, $\gamma$ a characteristic parameter of the randomness and the time integral is carried out along the considered path. This probability is a classical analog of the Feynman's factor $e^{\frac{i}{\hbar}A}$ of quantum Hamiltonian mechanics\cite{Feynman}, the difference being that $\gamma$ is a real number since the path probability must be real. The output of the numerical experiment has verified this theoretical prediction\cite{Wang1} and confirmed the generalization of the regular least action principle under these conditions.

A question following this work is whether or not the same exponential path probability exists for dissipative random motion. The answer to this question is crucial for the description of random motions in nature since most of them, if not all, are dissipative or over-damped. In this work, a analytical derivation of the path probability is provided first for the same dynamic model as in \cite{Wang1} in the zero dissipation limit and then for the case with important dissipation. The reader will find the above mentioned exponential path probability exists for both models. In the zero dissipation limit, the action is just the usual one as expected, while in the presence of important dissipation, the action is a new one suggested in a generalization of least action principle to dissipative motion\cite{Wang5}. This new action recovers the usual action when the dissipation tends to zero.

\section{The model}
\begin{figure} 
\includegraphics[width=5 in]{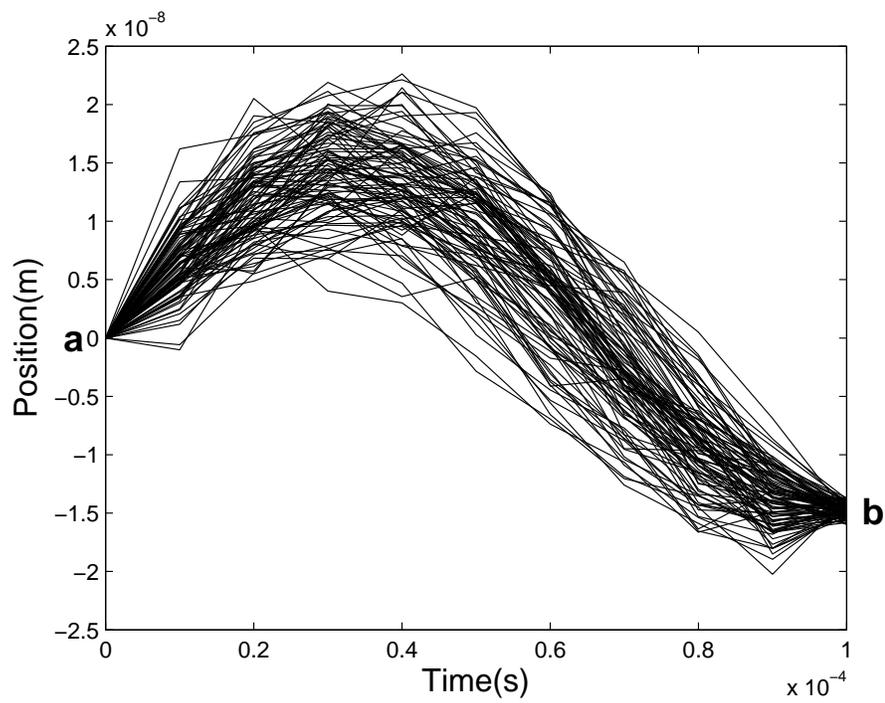}
\caption{Sample paths created with Eq.(\ref{2}) for a random motion subject to the Harmonic force and the Gaussian noise.}\label{fig.1}
\end{figure}

The model can be depicted as the following Langevin equation 
\begin{equation} \label{1}
m\frac{d^2x}{dt^2}=-\frac{dV(x)}{dx}+f_d+R
\end{equation}
where $x$ is the one dimensional position, $t$ the time, $V(x)$ the potential energy, $f_d$ the friction force, and $R$ the random forces as a white noise. A discrete solution of this equation is
\begin{equation} \label{2}
x_{i}=x_{i-1}+r_i+f(t_i)-f(t_{i-1})
\end{equation}
which is a superposition of a Gaussian distributed random displacement $r_i$ (solution of the stochastic equation $m\frac{d^2x}{dt^2}=R$) and a regular trajectory $f(t_i)$ (the solution of the Newtonian equation $m\frac{d^2x}{dt^2}=-\frac{dV(x)}{dx}+f_d$) corresponding to the least action path. Although Eq.(\ref{1}) is not necessarily a linear equation in $x$, Eq.(\ref{2}) is always its solution since the acceleration derivative $\frac{d^2}{dt^2}$ is a linear operator. Figure \ref{fig.1} illustrates some sample paths produced with Eq.(\ref{2}) between two positions $a$ and $b$, each of them being a tube of finite thickness $\delta$ (a band in the $x-t$ representation). 

In the numerical experiment of the work \cite{Wang1}, the number $N_k$ of the particles (or geometrical lines) going through a given sample path $k$, i.e., all the sequences of positions $\{x_{0},x_{1},x_{2}\cdots x_{n-1},x_{n}\}$ satisfying $\{z_{k,i}-\delta/2\leq x_{i}\leq z_{k,i}+\delta/2\}$ for all $i=1,2,\ldots,n$, is counted, where $\{a=z_{k,0},z_{k,1},z_{k,2},\ldots, z_{k,n-1},z_{k,n}=b\}$ are the axial line of that tube and $n$ the total number of steps. The probability that the path $k$ is taken is given by ${P_k=N_k/N}$ where $N$ is the total number of particles (trajectories) moving from $a$ to $b$ through all the considered sample paths. The discrete normalization of $P_k$ is $\sum_{k=1}^{s}P_k=1$ where $s$ is the total number of paths.

For each sample path, the instantaneous velocity of the particles of mass $m$ at the step $i$ is calculated by $v_{k,i}=\frac{z_{k,i}-z_{k,i-1}}{t_i-t_{i-1}}$ along the axial line. This velocity can be approximately considered as the average velocity of all the trajectories passing through the tube if $\delta$ is small with respect to the whole scale of the motion. The kinetic energy is given by $K_{k,i}=\frac{1}{2}mv_{k,i}^2$. The action along a path $k$ is given by $A_k=\sum_{i=1}^{n} [\frac{1}{2}mv_{k,i}^2-V(z_{k,i})]\cdot \Delta t$. 

The Gaussian distribution of $r_i$ at the step $i$ is given by

\begin{equation}   \label{3}
p(r_i,\sigma)=\frac{1}{\sqrt{2\pi}\sigma} e^{-\frac{r_i^2}{2\sigma^2}},
\end{equation}

\noindent where, in the case of a free Brownian particle for instance, $\sigma=\sqrt{2D(t_i-t_{i-1})}=\sqrt{2D\Delta t}$ is the standard deviation, $D$ the diffusion constant, and $\Delta t=t_i-t_{i-1}$ is the time interval of each step of the discrete motion.

\section{Path probability of random motion in the non dissipative limit}
This is the case where, in the Eq.(\ref{1}), the dissipative force $f_d$ is very weak with respect to the conservative force $-\frac{dV(x)}{dx}$, or the dissipative energy by the force $f_d$ is small compared to the variation of the total mechanical energy. 

From the definition of the Wiener measure\cite{Mazo}, the probability $P_k$ of a given path $k$ can be expressed as follows:

\begin{equation}    \label{4}
P_{k}=\frac{\prod_{i=1}^{n}p_{k,i}}{\sum_{k=1}^{s}\prod_{i=1}^{n}p_{k,i}}
\end{equation}

\noindent where $p_{k,i}=p(z_{k,i}-\delta/2\leq X_{i}\leq z_{k,i}+\delta/2)$ for all $i=1,2,\ldots,n$, is the Gaussian distribution of all the realization $x_i$ of the random variable $X_i$ that are found in the tube of the path $k$. From Eq.(\ref{2}), $p_{k,i}$ is given by

\begin{equation}     \label{5}
p_{k,i}=\frac{1}{\sqrt{2\pi}\sigma}\int_{z_{k,i}-\delta/2}^{z_{k,i}+\delta/2} \exp\{-\frac{[u-x_{i-1}-f(t_i)+f(t_{i-1})]^2}{2\sigma^2}\}du
\end{equation}

\noindent For small $\delta$, we have

\begin{equation}     \label{6}
p_{k,i}\approx\frac{\delta}{\sqrt{2\pi}\sigma}\exp\{-\frac{[z_{k,i}-x_{i-1}-f(t_i)+f(t_{i-1})]^2}{2\sigma^2}\}
\end{equation}

\noindent This leads to
\begin{equation}     \label{7}
P_k\approx\frac{1}{Z}\exp\{-\frac{1}{2\sigma^2}\sum_{i=1}^{n}[z_{k,i}-x_{i-1}-f(t_i)+f(t_{i-1})]^2\}
\end{equation}

\noindent where the normalization constant is given by
\begin{equation}     \label{8}
Z=\sum_{k=1}^s\exp\{-\frac{1}{2\sigma^2}\sum_{i=1}^{n}[z_{k,i}-x_{i-1}-f(t_i)+f(t_{i-1})]^2\}.
\end{equation}

\noindent This implies
\begin{equation}     \label{9}
\ln P_k\approx-\frac{1}{2\sigma^2}\sum_{i=1}^{n}[z_{k,i}-x_{i-1}-f(t_i)+f(t_{i-1})]^2-\ln Z.
\end{equation}
As $Z$ is constant, the function $\ln P_k$ depends only on the sum over $i$. Let it be $S_k$, we have

\begin{equation}    \label{10}
\begin{aligned}
S_k= \sum_{i=1}^{n}(f(t_i)&- f(t_{i-1}))^2+\sum_{i=1}^{n}(z_{k,i}-z_{k,i-1})^2+\sum_{i=1}^{n}(z_{k,i-1}-x_{i-1})^2\\&
+2\sum_{i=1}^{n}(z_{k,i}-z_{k,i-1})(z_{k,i-1}-x_{i-1})\\&+2\sum_{i=1}^{n}(z_{k,i-1}-x_{i-1})(f(t_i)-f(t_{i-1}))\\
&
-2\sum_{i=1}^{n}(z_{k,i}-z_{k,i-1})(f(t_i)-f(t_{i-1}))
\end{aligned}
\end{equation}

\noindent We remark that :

\noindent $\bullet$ The first term is constant because it is a sum over the deterministic path $f(t)$.

\noindent $\bullet$ The second term in Eq.(\ref{5}) is just the time sum (time integral in continuous version) of the kinetic energy, i.e., $\sum_{i=1}^{n}(z_{k,i}-z_{k,i-1})^2=\frac{2\Delta t}{m}\sum_{i=1}^{n}\frac{1}{2}mv_{k,i}^2\Delta t$.

\noindent $\bullet$ The third term $\sum_{i=1}^{n}(z_{k,i-1}-x_{i-1})^2\leq n\delta^2$ can be neglected for small $\delta$.

\noindent $\bullet$ The fourth term is such that 

\begin{equation}  \label{11}
2\sum_{i=1}^{n}(z_{k,i}-z_{k,i-1})(z_{k,i-1}-x_{i-1})\leq 2\delta n \max\limits_{\substack{k=1,\ldots,s\\i=1,\ldots,n}}\left|z_{k,i}-z_{k,i-1}\right|.
\end{equation}

\noindent It follows that this term can be neglected for small $\delta$ because $\max\limits_{\substack{k=1,\ldots,s\\i=1,\ldots,n}}\left|z_{k,i}-z_{k,i-1}\right|$ is constant.

\noindent $\bullet$ The fifth term is such that 
\begin{equation}  \label{12}
2\sum_{i=1}^{n}(z_{k,i-1}-x_{i-1})(f(t_i)-f(t_{i-1}))\leq 2\delta n \max_{\substack{i=1,\ldots,n}}\left|f(t_i)-f(t_{i-1})\right|.
\end{equation}

\noindent Consequently, this term can be neglected for small $\delta$ because $\max\limits_{\substack{i=1,\ldots,n}}\left|f(t_i)-f(t_{i-1})\right|$ is constant.

\noindent $\bullet$ Finally, the sixth term in Eq.(\ref{5}) reads

\begin{equation}   \label{13}
\begin{aligned}
-2\sum_{i=1}^{n}(z_{k,i}-z_{k,i-1}&)(f(t_i)-f(t_{i-1}))=2\sum_ {i=1}^{n-1}z_{k,i}\left(f(t_{i-1}-2f(t_i)+f(t_{i+1})\right)\\
+&2z_{k,0}[f(t_1)-f(t_0)]+2z_{k,n}[f(t_n)-f(t_{n-1})]\\
=&2\sum_ {i=1}^{n-1}z_{k,i}({\Delta t})^2\left(\frac{\frac{f(t_{i+1})-f(t_i)}{\Delta t}-\frac{f(t_{i})-f(t_{i-1})}{\Delta t}}{\Delta t}\right)\\
+&2a[f(t_1)-f(t_0)]+2b[f(t_n)-f(t_{n-1})]\\
\end{aligned}
\end{equation}

\noindent It follows that for $({\Delta t})$ small, we have

\begin{equation}    \label{14}
\begin{aligned}
-2\sum_{i=1}^{n}(z_{k,i}-z_{k,i-1})(f(t_i)-&f(t_{i-1}))
\approx 2\sum_ {i=1}^{n-1}z_{k,i}({\Delta t})^2\left(\ddot{f}(t_i)\right)\\
+&2a[f(t_1)-f(t_0)]+2b[f(t_n)-f(t_{n-1})]\\
\end{aligned}
\end{equation}

\noindent At this stage, $S_k$ reads

\begin{equation}    \label{15}
S_k=\frac{2\Delta t}{m}\left[\sum_{i=1}^{n}\frac{1}{2}mv_{k,i}^2\Delta t+{\Delta t}\sum_ {i=1}^{n-1}z_{k,i}m\ddot{f}(t_i)\right]+constant
\end{equation}

\noindent where $constant$ includes all the constant terms independent of $k$ mentioned above. Since $m\ddot{f}(t_i)=-\frac{dV(f(t_i))}{df(t_i)}=-V'(f(t_i))$, we get

\begin{equation}    \label{16}
S_k=\frac{2\Delta t}{m}\left[\sum_{i=1}^{n}\frac{1}{2}mv_{k,i}^2\Delta t-\sum_ {i=1}^{n-1}z_{k,i}V'(f(t_i)){\Delta t}\right]
\end{equation}
to a constant. In order to introduce the potential energy $V(z_{k,i})$ on the path $k$, we consider the following approximation
\begin{equation}    \label{17}
V(z_{k,i})\approx V(f(t_i))+(z_{k,i}-f(t_i))V'(f(t_i)),
\end{equation}

\noindent or $z_{k,i}V'(f(t_i))\approx V(z_{k,i})-V(f(t_i))+f(t_i))V'(f(t_i))$, which leads to

\begin{equation}    \label{18}
S_k=\frac{2\Delta t}{m}\left[\sum_{i=1}^{n}\frac{1}{2}mv_{k,i}^2\Delta t-\sum_ {i=1}^{n-1}V(z_{k,i}){\Delta t}\right]
\end{equation}

\noindent to a constant including all the terms independent of of the path $k$ in the Eq.(\ref{17}). Let us introduce the usual action of the mechanics $A_k$ given by

\begin{equation}    \label{19}
A_k=\sum_{i=1}^{n}L_{k,i}\Delta t
\end{equation}

\noindent where the Lagrangian is ${L_{k,i}=K_{k,i}-V(z_{k,i})}$, the kinetic energy  ${K_{k,i}=\frac{1}{2}mv_{k,i}^2}$, and the potential energy $V(z_{k,i})$, all along the path $k$, up to the time $t_i$. Finally, the probability of the path $k$ is given by 
\begin{equation}    \label{20}
P_k=\frac{1}{Z}\exp(-\eta A_k)
\end{equation}
with $\eta=\frac{\Delta t}{m\sigma^2}$. All the terms independent of $k$ vanish after the normalization over $k$ with the constant ${Z=\sum_{k=1}^s\exp(-\eta A_k)}$. As $\eta$ is always positive, Eq.(\ref{20}) means that the most probable path must be the paths of least action $A_k$. This probability has been observed in the numerical experiment of random motion in the weak dissipative limit in the work \cite{Wang1}. 

\section{Path probability of dissipative random motion}
The usual action loses its role of characteristic variable for the paths in the variational calculus when dissipation is present. The search for an alternative variable has been since long a headache for physicists in analytical mechanics, as discussed briefly in \cite{Wang5}.

In order to introduce the dissipative force $f_d$ and the energy it dissipated in a Lagrangian to calculate the action, we consider the work $E_d$ of $f_d$, along the path $f(t_i)$, from the time $t=t_0=0$ to the time $t=t_i$, i.e.,  $E_d(f(t_i))=-\int_0^{f(t_i)}f_d(x)dx$. $E_d$, being the energy lost by the system to the environment, must be positive. We have, according to the second fundamental theorem of calculus, the following expression $f_d=-\frac{dE_d(f(t_i))}{df(t_i)}=-E_d'(f(t_i))$. 

Now Eq.(\ref{16}) can be rewritten with the help of the following equation of motion 
\begin{equation}    \label{21}
m\ddot{f}(t_i)=-V'(f(t_i)-E_d'(f(t_i))
\end{equation}
and the approximation
\begin{equation}    \label{22}
\begin{aligned}
V(z_{k,i})+E_d(z_{k,i})&\approx V(f(t_i))+E_d(f(t_i))\\
&+(z_{k,i}-f(t_i))\left[V'(f(t_i))+E_d'(f(t_i))\right]
\end{aligned}
\end{equation}
\noindent After arrangement, $S_k$ reads, to a constant independent of the index $k$, 
\begin{equation}    \label{23}
S_k=\frac{2\Delta t}{m}A_k
\end{equation}
where the action $A_k$ is given by
\begin{equation}    \label{24}
A_k=\sum_{i=1}^{n}L_{k,i}\Delta t
\end{equation}
with the Lagrangian ${L_{k,i}=K_{k,i}-V(z_{k,i})-E_d(z_{k,i})}$, the kinetic energy is  ${K_{k,i}=\frac{1}{2}mv_{k,i}^2}$, the conservative potential energy $V(z_{k,i})$ and the energy dissipated $E_d(z_{k,i})$, along the path $k$ up to the time $t_i$. Finally, the path probability of the path $k$ is given by 
\begin{equation}    \label{25}
P_k=\frac{1}{Z}\exp(-\eta A_k)
\end{equation}
All the terms independent of $k$ vanish after the normalization over $k$ with the constant ${Z=\sum_{k=1}^s\exp(-\eta A_k)}$. 

The quantity $A_k$ is still called action for dissipative system because, in a variational calculus proposed in \cite{Wang5}, its vanishing variation $\delta A_k=0$ can lead to the Newtonian equation of motion with dissipative force. It is then a characteristic variable of paths for dissipative motion in the same way as the usual action for non dissipative motion of Hamiltonian systems. This `dissipative' action recovers the usual action for the case where $E_d$ is negligible with respect to the energy of the system.

\section{Concluding remarks}
In this work, the path probability for random dynamics of mechanical system has been calculated on the basis of a model described by Langevin equation with Gaussian noise. The result shows an exponential dependence of the probability on the action which confirms the measure of the path probability by numerical simulation of random motion with the same model in the non dissipative limit. In the case of dissipative motion, the action can be calculated with a Lagrangian which include the energy lost by the system due to the dissipation, which confirms a generalization of the principle of least action of Hamiltonian systems to dissipative systems\cite{Wang5}. According to the result of this work, for both non dissipative and dissipative systems, the exponential path probability implies that the most probable paths are the paths of least action which, in the limit of vanishing randomness, become the regular paths of least action.

\end{document}